\documentclass[useAMS,usenatbib]{mn2e}

\usepackage{times}
\usepackage{graphicx}
\usepackage{amsmath}
\usepackage{amssymb}
\usepackage{color}
\usepackage{epsfig}
\usepackage{psfig}
\usepackage{journal_names}
\usepackage{paralist}

\newcommand{\kms}{\mbox{\,km~s$^{-1}$}}
\title[Mass modelling of NGC~1407]
{The SLUGGS Survey: Multi-population dynamical modelling of the elliptical galaxy NGC~1407 from stars and globular clusters}
\author[Pota~et~al.~ ]
{Vincenzo Pota$^{1,2}$, Aaron J. Romanowsky$^{2,3}$,  Jean P. Brodie$^2$, Jorge Pe\~narrubia$^4$,\\
\\
\normalfont{\LARGE Duncan A. Forbes$^{1}$, Nicola R. Napolitano$^5$, Caroline Foster$^6$, Matthew G. Walker$^8$} \\ 
\\
\normalfont{\LARGE Jay Strader$^7$, Joel C. Roediger$^9$} \\ 
\\
$^1$ Centre for Astrophysics \& Supercomputing, Swinburne University, Hawthorn VIC 3122, Australia\\
$^2$ University of California Observatories, 1156 High Street, Santa Cruz, CA 95064, USA\\
$^3$ Department of Physics and Astronomy, San Jos\'e State University, One Washington Square, San Jose, CA 95192, USA\\
$^4$ Institute for Astronomy, University of Edinburgh, Royal Observatory, Blackford Hill, Edinburgh EH9 3HJ, UK\\
$^5$ INAF-Observatory of Capodimonte, Salita Moiariello, 16, 80131, Naples, Italy\\
$^6$ Australian Astronomical Observatory, PO Box 915, North Ryde, NSW 1670, Australia\\
$^7$ Department of Physics and Astronomy, Michigan State University, East Lansing, Michigan 48824, USA\\
$^8$ McWilliams Center for Cosmology, Department of Physics, Carnegie Mellon University, 5000 Forbes Ave., Pittsburgh, PA 15213, USA \\
$^9$ NRC Herzberg Astronomy \& Astrophysics; Victoria, BC, V9E 2E7, Canada}

\begin{document}

\label{firstpage}

\maketitle
\begin{abstract}

We perform in-depth dynamical modelling of the luminous and dark matter (DM) content of the elliptical galaxy NGC~1407. Our strategy consists of solving the spherical Jeans equations for three independent dynamical tracers: stars, blue GCs and red GCs in a self-consistent manner. 
We adopt a maximum-likelihood Markov-Chain Monte Carlo fitting technique in the attempt to constrain the inner slope of the DM density profile (the cusp/core problem), and the stellar initial mass function (IMF) of the galaxy. 
We find the inner logarithmic slope of the DM density profiles to be $\gamma = 0.6\pm0.4$, which is consistent with either a DM cusp ($\gamma = 1$) or with a DM core $(\gamma = 0)$. Our findings are consistent with a Salpeter IMF, and marginally consistent with a Kroupa IMF. We infer tangential orbits for the blue GCs, and radial anisotropy for red GCs and stars.
The modelling results are consistent with the virial mass--concentration relation predicted by $\Lambda$CDM simulations. The virial mass of NGC~1407 is $\log M_{\rm vir} = 13.3 \pm 0.2 M_{\sun}$, whereas the stellar mass is $\log M_* = 11.8 \pm 0.1 M_{\sun}$. 
The overall uncertainties on the mass of NGC~1407 are only 5 per cent at the projected stellar effective radius. We attribute the disagreement between our results and previous X-ray results to the gas not being in hydrostatic equilibrium in the central regions of the galaxy. 
The halo of NGC~1407 is found be DM dominated, with a dynamical mass-to-light ratio of $M/L=260_{-100} ^{+174} M_{\sun}/L_{\sun, B}$. However, this value can be larger up to a factor of 3 depending on the assumed prior on the DM scale radius.
 
\end{abstract}


\section{Introduction}

The idea that the Universe is dominated by cold, non interacting dark matter (DM) and by dark energy ($\Lambda$) has been extensively tested over the last two decades. Computer simulations make specific predictions about the properties of DM haloes at $z=0$, some of which can be tested with the increasing quality of observational data. The emerging picture is that, although the $\Lambda$CDM model can generally explain observables on large scales, it may fail when it comes to galactic or sub-galactic scales \citep[e.g.,][]{Weinberg13,Kroupa14}. 


The modelling of DM in elliptical galaxies is notoriously difficult when compared to dwarf galaxies or spiral galaxies 
\citep{Walter08,Herrmann09,Kuzio08,Battaglia08,Amorisco12,Cole12,Amorisco13,Adams14}. Unlike dwarf spheroids, ellipticals are baryon dominated at the very centre. Our ignorance about the initial mass function (IMF) in ellipticals introduces a degeneracy between the DM mass and the stellar mass, which limits our ability to test $\Lambda$CDM predictions on galactic scales. Moreover, unlike spiral galaxies whose haloes are traced by HI gas \citep[e.g.,][]{Oh11}, the outer regions of elliptical galaxies are notoriously hard to observe because they are optically faint. Nevertheless, they can be probed with discrete tracers, such as planetary nebulae \citep[e.g.,][]{deLorenzi08,Napolitano}, globular clusters (GCs) \citep[e.g.,][]{DeasonB,Schuberth12}, or diffuse tracers, such as hot X-ray gas \citep[e.g.,][]{Humphrey06}. Lastly, some mass modelling techniques suffer from the mass-anisotropy degeneracy, driven by the fact that the orbital anisotropy of stellar systems is very hard to infer from the data \citep{Mamon05}.

It has been shown that the cumulative effect of model degeneracies on the model outcome can be alleviated by modelling of multiple dynamical tracers within the same galaxy \citep[e.g.,][]{Walker,Schuberth,NewmanB,Napolitano14}. This approach can drastically reduce modelling uncertainties by up to a factor of four.

In this paper, we model the luminous and DM content of the elliptical galaxy NGC~1407 using three independent dynamical tracers: the stars, which probe the innermost effective radius of the galaxy $(R_e)$, and GCs, which in turn consist of two independent (red and blue) subpopulations, out to 10 $R_e$. The blue and the red GCs are thought to represent different stages of galaxy evolution \citep{Brodie} and are characterized by distinct kinematic and spatial properties \citep[e.g.,][]{Puzia04,Lee10,Woodley10,Norris12,Pota13}. Our strategy is to build a self-consistent model with minimal assumptions, in the attempt to constrain DM parameters that are directly comparable to $\Lambda$CDM predictions.

Simultaneous modelling of stars and GCs has been carried out in other elliptical galaxies \citep{Romanowsky01,Schuberth,Schuberth12,Napolitano14}. For one galaxy (M~87, \citealt{Agnello14B}), it has been shown that this method can differentiate between cuspy and cored DM profiles. The flattening, or steepening, of the DM inner slope is currently very relevant for galaxy evolution theories, because it can be related to a number of competing physical processes, such as DM contraction \citep{Gnedin04}, baryonic feedback \citep{Cole11,DiCintio13,Martizzi13}, or self-interacting DM \citep{Rocha13,Peter13}. 

Our choice of NGC~1407 is also motivated by the rich dataset collected as part of the SAGES Legacy Unifying Globulars and GalaxieS (SLUGGS\footnote{http://sluggs.ucolick.org}) survey \citep{Brodie14}, and by previous claims that NGC~1407 has a very elevated $B$-band mass-to-light ratio $M/L\approx800 M_{\sun} / L_{\sun}$ \citep[e.g.,][]{Gould93}, in tension with scaling relations for galaxy groups \citep{Eke06}. On the other hand, modelling of X-ray gas in this galaxy \citep{Humphrey06,Zhang,Su14} found values of $M/L$ that are lower by a factor of two. 

This paper is structured as follows. The dataset and the observables needed to solve the Jeans equations are discussed in Section \ref{sec:data}.  The dynamical model and the fitting method are both given in Section \ref{sec:Jeans}. The modelling results from Section \ref{sec:results} are discussed in Section \ref{sec:discussion}, where we compare our findings with literature studies, computer simulations and mass estimators. Section \ref{sec:summary} summarises the results of the paper.
 
\section{Data}
\label{sec:data}

In this section we discuss the derivation of the observables needed to solve the Jeans equations in Section \ref{sec:Jeans}. We are interested in the one-dimensional radial distribution of stars and GCs, along with the velocity dispersion profile for both tracers. The data are products the SLUGGS survey and have been partially discussed in a series of papers \citep{Proctor,Foster09,Spitler12,Pota13,Arnold14}. 

\subsection{Target and conventions} 
\label{sec:distance}

NGC 1407 is a massive E0 galaxy at the centre of the dwarf galaxy dominated Eridanus A group \citep{Brough}. It shows moderate rotation along the photometric major axis and evidence of past AGN activity \citep{Giacintucci12}. In this paper, we derive an absolute magnitude of $M_B=-21.84$ mag and a total luminosity of $L_B = 8.53\times 10^{10} L_{\sun}$ (see \S \ref{sec:SB}). The galaxy appears relaxed in optical imaging, but it shows disturbances both in X-ray imaging and in velocity space, as we will discuss throughout the paper. The galaxy minor-to-major axis ratio is $q=0.95$ and it does not show any significant variation with radius out to 280 arcsec \citep{SpolaorB,Li11}. The adopted position angle and systemic velocity are PA$=$55 deg \citep{SpolaorB} and $v_{\rm sys} = 1779 \kms$ (from NED).

The distance to NGC~1407 is very uncertain. Using the GC luminosity function, \citet{Forbes06A} found $D=20.9$ Mpc, whereas surface brightness fluctuations return distances between 25 Mpc to 29 Mpc \citep{Tonry01,Cantiello05} . 
In this paper we will assume $D=28.05$ Mpc as in \citet{Rusli2013a}. The distance modulus is therefore $(m-M)=32.24$ mag. With this distance assumption, 1~arcsec corresponds to 0.136 kpc. 

Throughout the paper, we will refer to the galactocentric 2D (projected) radius and 3D (de-projected) radius as $R$ and $r$, respectively. The galactocentric radius is defined as the circularized radius $R^2=(q \, X^2) + (Y^2/q)$, where $X$ and $Y$ are aligned with the galaxy major axis and minor axis, respectively. All quoted magnitudes were reddening corrected according to the dust map of \citet{Schlegel}.

\begin{figure}
\includegraphics[width=\columnwidth]{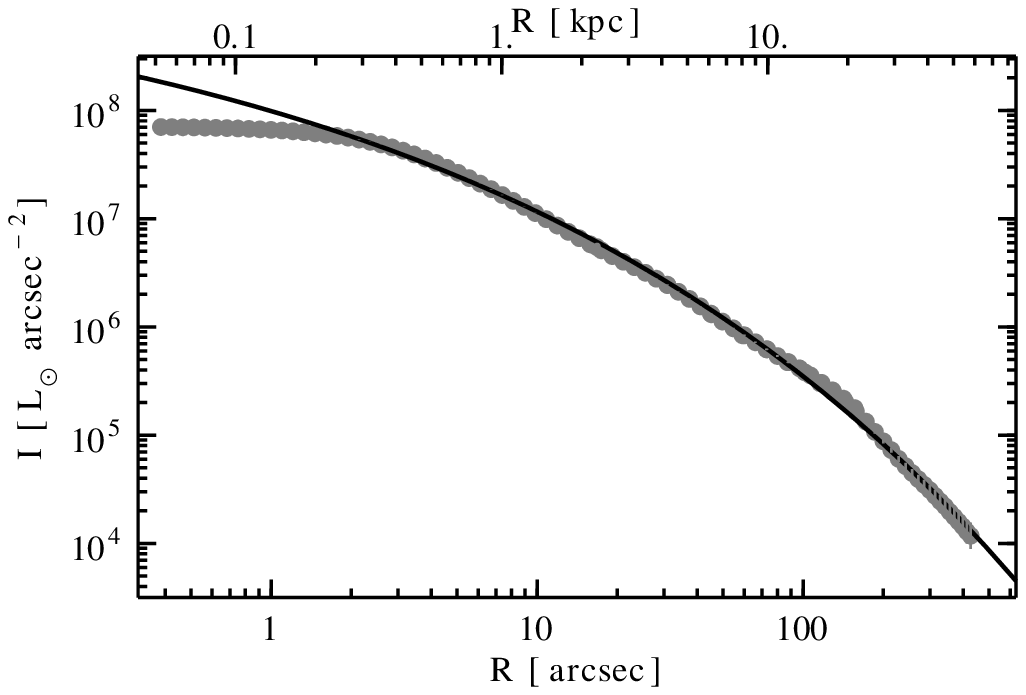} 
\includegraphics[width=\columnwidth]{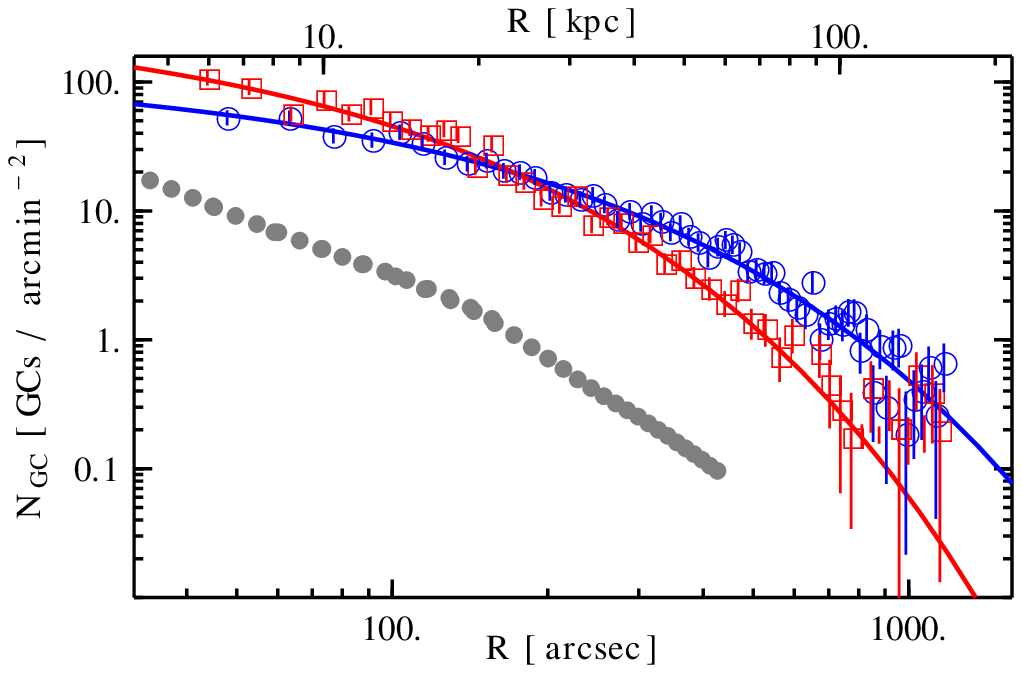} 
\caption{Radial surface density profiles of our dynamical tracers. Physical scales are given on the top axes. The top panel shows the $B$-band stellar surface brightness data points (grey), and the best fit S\`ersic law to the data (black line). We do not fit the inner core. The bottom panel shows surface density data points for blue (blue circles) and red GCs (red squares) along with the S\`ersic best fits. The grey points are the stellar surface brightness data arbitrarily rescaled for comparison purposes. Overall, the three dynamical tracers have very different density distributions.}
\label{fig:I(R)}
\end{figure}

\subsection{Stellar surface brightness}
\label{sec:SB}

We combined the $B$-band surface brightness profile from HST/ACS \citep{SpolaorB,Rusli2013b}, with Subaru/Suprime-Cam data in the $g$-band presented in \citet{Pota13}.

The Subaru $g$-band surface brightness profile was derived with the XVista software package\footnote{http://astronomy.nmsu.edu/holtz/xvista/index.html}. Bright sources and the nearby galaxy NGC~1400 were masked out in order to maximize the signal from NGC~1407. The $g$-band surface brightness profile was renormalized to match the $B$-band profile because these two filters have very similar passbands. The wide-field of Suprime-Cam (34 $\times$ 27 arcmin$^2$) allows us to measure the galaxy surface brightness out to 440 arcsec from the centre. 
We convert $B$ magnitudes to solar luminosities by adopting a solar absolute magnitude of $M_{\sun , B}=5.48$ mag. The resulting luminosity profile is shown in Figure \ref{fig:I(R)}. We can see that NGC~1407 has a central stellar core and a bumpy luminosity profile outside 100 arcsec. These features can be fitted via multiple \citet{Sersic} profiles, as performed by \citet{Li11} and \citet{Rusli2013a}. 

For our current purposes, we perform a single S{\'e}rsic fit to the data points in Figure \ref{fig:I(R)}, after masking the innermost 2 arcsec. We find a S{\'e}rsic index $n=4.67 \pm 0.15$, an effective radius $R_e=100 \pm 3 \mbox{ arcsec} = 13.6 \pm 0.4$ kpc, and an intensity $I(R~=~R_e)~=~3.5~\times~10^{5} L_{\sun}  \mbox{ arcsec}^{-2}$. From the best fit to the data, we infer a $B$-band absolute magnitude of $M_B=-21.84$ mag and a total luminosity of $L_B = 8.53\times 10^{10} L_{\sun}$. 

We note that our estimate of $R_e$ is a factor of two larger compared to values from the literature \citep[e.g.,][]{SpolaorB}, but it is consistent with size-luminosity scaling relations from wide-field observations \citep{Kormendy09}. Such large values of $R_e$ have become increasingly common with the advent wide-field photometry, which is able to detect faint surface brightness features in galaxy haloes \citep[e.g.,][]{Donzelli11}. 

\subsection{Globular cluster surface density}
\label{sec:surfacedensity}

We extracted our GC catalogue from Subaru/Suprime-Cam imaging in $gri$ filters. Details on the data acquisition and reduction are given in \citet{Pota13}. Briefly, GC candidates were selected in colour-colour and colour-magnitude space. The resulting GC colour distribution is clearly bimodal, with a dividing colour between blue and red GCs at $(g-i) \approx 0.98$ mag. This result is in agreement with an HST-based study of the NGC~1407 GC system \citep{Forbes06A}.

The GC surface density was obtained by binning candidates brighter than $i<25.5$ mag in circular annuli centered on the galaxy, and dividing the resulting number by the area of the corresponding annulus. Each annulus contained more than 50 GCs per bin. Poissonian errors on the surface density were calculated as the ratio $\sqrt{\mbox{GCs per bin}} / \mbox{Area of the annulus}$. 

The resulting (blue and red) GC surface density profiles were fitted with a modified S{\'e}rsic function of the form:
\begin{equation}
N_{\rm GC}(R) = N_e \times \exp \left\lbrace -b_n \left[ \left( \frac{R}{R_e} \right)^{1/n} -1 \right] \right\rbrace + bg,
\label{eq:NGC}
\end{equation}
where $N_e$ is the numerical surface density at $R_e$, $b_n= 2 n - 1/3 + 009876 n^{-1}$ \citep{Ciotti99}, and $bg$ is the background contamination level, which was assumed to be homogeneous across the image. 

The best fit to the GC surface density profiles of blue and red GCs are shown in Figure \ref{fig:I(R)}. The best fit values are given in Table \ref{tab:sersic}. The red GCs are more centrally concentrated than the blue GCs, as found in other galaxies \citep[e.g., ][]{Forte,Bassino,Strader11,Forbes12}. 

\begin{figure}
\includegraphics[width=\columnwidth]{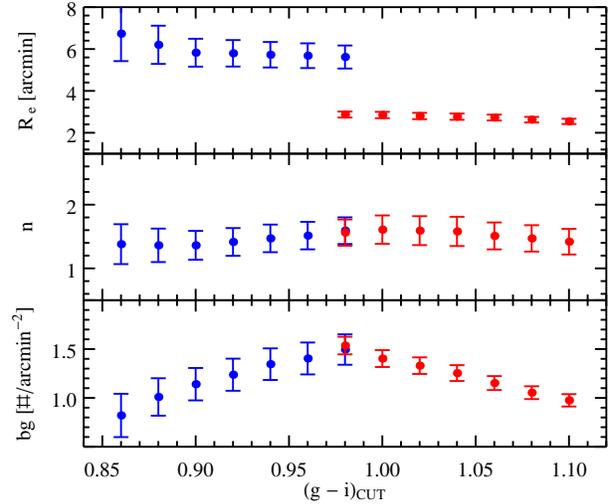} 
\caption{Photometric GC properties as a function of $(g-i)$ colour. From the top to the bottom we show the projected effective radius $R_e$, the S\'ersic index $n$ and the background level $bg$ computed for different colour cuts at $(g-i)_{\rm CUT}$.}
\label{fig:colourcut}
\end{figure}

\begin{table}
\centering
\label{mathmode}
\begin{tabular}{@{}l l c c c}
\hline
Sample & $R_e$ & $n$ & $N_e$ & $bg$\\
&  [kpc] & & [GCs  arcmin$^{-2}$] & [GCs  arcmin$^{-2}$] \\ \hline
Stars & $13.6\pm0.4$ & $4.67\pm0.15$ & -- & -- \\
Blue GCs &$47\pm4$ & 1.6$\pm$0.2 & 7$\pm$1&1.4$\pm$0.2 \\ 
Red  GCs &$23\pm1$ & 1.6$\pm$0.2 & 20$\pm2$&1.4$\pm$0.1 \\
\hline
\end{tabular}
\caption{Density profile best fit parameters for stars and GCs. A S{\'e}rsic function was fitted to both stars and GCs. For the latter, a colour cut at $(g~-~i)= 0.98$ was adopted to separate blue and red GCs.}
\label{tab:sersic} 
\end{table}

\subsection{The effect of the colour cut}
\label{sec:fiducial}

We investigate the effect of the adopted colour cut on the inferred parameters ($R_e$ and $n$ in particular) from eq.\ref{eq:NGC}. We select the bluest and reddest GCs with an initial colour cut at $(g-i)=0.98$ mag, and then we slide the colour cut toward the blue and red wings of the colour distribution, thereby clipping the GCs with intermediate colour. We fit eq. \ref{eq:NGC} for each adopted colour cut and we study how the photometric parameters vary with this quantity. 

The results are shown in Figure \ref{fig:colourcut}, where the $x$-axis represents the adopted colour cut. 
We can see that the best fit parameters to eq. \ref{eq:NGC} are remarkably constant regardless of the colour cut. The S\`ersic index is stable at $n\approx1.5\pm0.2$, whereas the effective radii of the two subpopulations are different, as found in almost all GC systems studied so far. 

We experimented with different bin sizes, finding that 30 to 80 GCs per bin return results consistent with those in Figure \ref{fig:colourcut}, where 60 GCs per bin was adopted. The scatter introduced by varying the bin size across this range is included in the uncertainties in Figure \ref{fig:colourcut}. 
In light of these results, we decided to perform our mass modelling with our fiducial colour cut set at $(g~-~i)_{\rm CUT}= 0.98$ mag. This cut was applied both to the spectroscopic and to the photometric sample. 

\subsection{Stellar velocity dispersion}
\label{sec:stellarVD}

We combine stellar kinematic data from two telescopes. We use ESO 3.6/EFOSC2 major axis long-slit data for the radial range between 0 and 40 arcsec from \citet{Proctor}. We also use multi slit Keck/DEIMOS data from Foster et al. (2015) for the radial range between 30 and 110 arcsec.

The root-mean-square velocity dispersion $v_{\rm rms}$ profile was obtained by folding and averaging the data with respect to the galaxy center\footnote{Following \citet{Napolitano}, we estimate the $v_{\rm rms}$ of long slit data as $v_{\rm rms}^2 \approx v_{\rm rot}^2 /2 + \sigma^2$ because we have data only along the major axis. This has only a small impact on the final result, given that $v_{\rm rot}\ll \sigma$ at all radii.}. Given the lack of apparent rotation in this galaxy (Foster et al. 2015), the $v_{\rm rms}$ is equivalent to the classic velocity dispersion $\sigma$. We will use the quantity $v_{\rm rms}$ throughout the paper.

\begin{figure}
\includegraphics[width=\columnwidth]{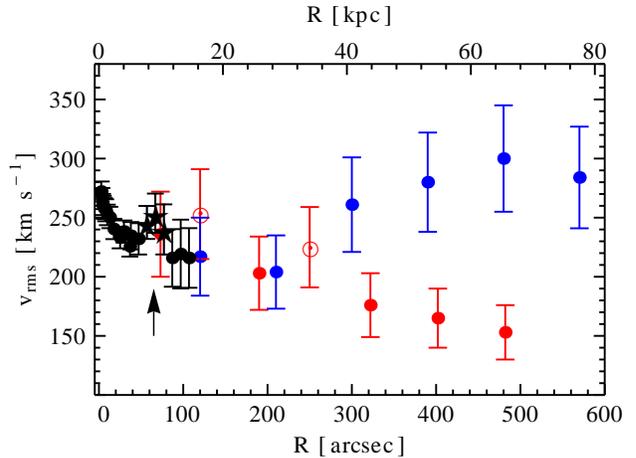} 
\caption{Velocity dispersion profile for our dynamical tracers. The $v_{\rm rms}$ profile (defined in \S \ref{sec:GCVD} and \ref{sec:stellarVD}) for stars, blue GCs and red GCs is shown in black, blue and red, respectively. Open red points mark the radial bins at which velocity dispersion bumps are detected. The black arrow localizes the velocity dispersion bump of the stars, which is also shown by black star symbols. Blue and red GCs have very different velocity dispersion profiles outside 200 arcsec, whereas red GCs and stars are alike.}
\label{fig:vrms}
\end{figure}

The stellar velocity dispersion profile of NGC~1407 is shown in the innermost 100 arcsec of Figure \ref{fig:vrms}. The $v_{\rm rms}$ profile declines smoothly with radius, and it shows a velocity dispersion bump ($\sigma$-bump) between $50<R<90$ arcsec, which is also detected in metallicity space \citep{Pastorello14}. We will discuss how this feature affects our results further in the text. 

\subsection{Globular cluster velocity dispersion}
\label{sec:GCVD}

The spectroscopic GC sample was derived from the 9 DEIMOS masks discussed in \citet{Pota13}, plus one additional DEIMOS mask observed on 2013 September 29. The new observations consisted of 4 exposures of 1800 seconds each. The seeing was 0.73 arcsec. The instrument configuration, data reduction and data analysis were identical to those described in \citet{Pota13}. From this additional mask, we acquired 23 newly confirmed GCs and 16 GCs duplicates from previous masks. Overall, our spectroscopic sample consists of 379 confirmed GCs. 

From this catalogue, we clip 6 GCs with radial velocities deviating more than $3\sigma$ from the velocity distribution of the $N=20$ closest neighbours \citep{Merrett}. This is performed for the blue and red GCs separately because these two subpopulations have very different velocity dispersion profiles. 
We also clip 72 GCs brighter than $\omega$ Cen, the brightest star cluster in the Milky Way, which we set as the boundary between GCs and ultra compact dwarfs (UCDs). UCDs were excluded from our analysis because they represent a tracer population which is kinematically and spatially distinct from normal GCs \citep[e.g.,][]{Zhang15,Strader11}. For $\omega$ Cen, we assume a magnitude $M_i = -11.0$ mag ($i=21.2$ mag). Our final GC catalogue consists of 153 blue GCs and 148 red GCs. 
 
We calculate the root mean square velocity dispersion of the two GC subpopulations as:
\begin{equation}
v_{\rm rms}^2 = \frac{1}{N}  \sum (v_i - v_{\rm sys})^2 - (\Delta v_i)^2 .
\label{eq:sigma}
\end{equation}
where $v_i$ is the radial velocity of the i-th GC and $\Delta v_i$ is its uncertainty. $v_{\rm sys}=1779 \kms$ is the systemic velocity of NGC~1407. Uncertainties were derived with the formulae provided by \citet{Danese}. We solve eq. \ref{eq:sigma} for radial bins with 25-30 GCs per bin.

Figure \ref{fig:vrms} shows that the velocity dispersion of the blue GCs increases with radius, whereas that of the red GCs decreases with radius. The velocity dispersion of the red GCs shows two bumps at $R \approx 150$ arcsec and $R \approx 250$ arcsec respectively, as found in the stellar kinematic profile, although at a different galactocentric radius. The contribution of GC rotation is naturally folded into eq. \ref{eq:sigma}, although we find that rotation is small, $(v_{\rm rot}/ \sigma) < 0.4$ for both GC subpopulations \citep{Pota13}.

\section{Mass modelling}
\label{sec:Jeans}

Our dynamical model is based on the study of the velocity moments of stars and GCs around NGC~1407. Apart from spherical symmetry and dynamical equilibrium, we also assume that all dynamical tracers are pressure supported, as suggested by the apparent lack of rotation both in the stars and GCs. In this context (i.e., small $v_{\rm rot}$), the projected velocity dispersion $\sigma$ is very similar to the root-mean-square velocity dispersion $v_{\rm rms} ^2 = \sigma^2 + v_{\rm rot}^2$.

\subsection{Method}

We want to relate the observed velocity dispersion profiles $v_{\rm rms}(R)$, and the density profiles $I(R)$ of our dynamical tracers, to the total mass in NGC~1407. This can be achieved by solving the spherical \citet{Jeans15} equations, with a solution of the form \citep{Mamon05}:
\begin{equation}
\label{eq:vrms}
v_{\rm rms} ^2 (R) = \frac{2 G}{I(R)} \int_{R}^{\infty} 
 K \left(\beta, \frac{r}{R} \right) \nu(r) M(r) \frac{{\rm d} r}{r} ,
\end{equation}
where $I (R)$ is the 2D (projected) density profile of the tracer, $\nu(r)$ is the 3D (de-projected) density profile and $M$ is the total mass of the galaxy within the radius $r$. The parameter $\beta = 1 - \sigma^2 _{\theta} / \sigma^2 _r$ is the orbital anisotropy which represents the ratio of the velocity dispersion in the tangential direction $\sigma^2 _{\theta}$ to the velocity dispersion in the radial direction $\sigma^2 _r$.
We assume the anisotropy to be constant with radius. The parameter $K$ is a complicated function of the anisotropy $\beta$ (equation A16 from \citealt{Mamon05}).
We obtain $\nu(r)$ by de-projecting the observed luminosity profile of the stars $I(R)$, and the observed number density for the GCs $N_{\rm GC}(R)$, respectively. The de-projection is performed via the numerical approximation of \citet{Prugniel}. 

The total mass of the galaxy is:
\begin{equation}
M(r) = M_* (r) +  M_{\rm d} (r),
\label{eq:Mtot}
\end{equation}
where $M_*$ and $M_{\rm d}$ are the mass of the stellar and the DM component, respectively.

The stellar mass is obtained by integrating the de-projected stellar luminosity profile $\nu_*$:
\begin{equation}
M_* (r) = 4 \, \pi \Upsilon_* \int_0^r \nu_* (r) r^2 \mathrm{d}r ,
\label{eq:masses}
\end{equation}
where $\Upsilon_* \equiv (M/L)_*$ is the stellar mass-to-light ratio, assumed to be constant throughout the galaxy. 

We parametrize the DM density profile with a generalized Hernquist profile \citep{Hernquist90,Zhao96}, also dubbed the generalized Navarro-Frenk-White profile (gNFW):
\begin{equation}
\rho_d (r) =  \rho_s \left( \frac{r}{r_s} \right)^{-\gamma} \left[ 1+ \left( \frac{r}{r_s} \right) \right]^{\gamma - 3}
\label{eq:Hern}
\end{equation}
where $r_s$ and $\rho_s$ are the characteristic DM scale radius and DM density, respectively. When $r \ll r_s$, the DM profile declines as $r^{-\gamma}$. When $\gamma=1$, the DM has a central cusp \citep{Navarro97}, whereas when $\gamma=0$ the DM has a central DM core.
When $r \gg r_s$ the DM profile declines as $r^{-3}$, as found for relaxed DM haloes at $z=0$ in $\Lambda$CDM simulations \citep{Navarro97}. The integration of eq. \ref{eq:Hern} gives the cumulative DM mass within a given radius:
\begin{equation}
M_d (r) =  \frac{4 \pi \rho_s r_s ^3}{\omega} \left( \frac{r}{r_s} \right)^{\omega} \,  _2F_1 \left[ \omega, \omega ; \omega ; -\left( \frac{r}{r_s} \right) \right]
\label{eq:Mhern}
\end{equation}
where $\omega=3-\gamma$ and $_2F_1$ is Gauss' hypergeometric function. 

\subsection{Maximum likelihood analysis}
\label{sec:maximum}

We want to find a set of model parameters able to simultaneously reproduce (via eq. \ref{eq:vrms}) the empirical velocity dispersion profiles of stars, blue GCs and red GCs. This can be achieved by considering that the three dynamical tracers are embedded in the same gravitational potential (stars$+$DM halo). In our model, the galaxy potential is characterized by the parameters: $r_s$, $\rho_s$, $\gamma$ (which characterize the DM halo) and by $\Upsilon_*$ (which characterizes the stellar mass). On the other hand, the orbital anisotropy of stars, blue GCs and red GCs (namely $\beta_*$,$\beta_B$,$\beta_R$, respectively) is a physical quantity uniquely associated with each dynamical tracer. 

We adopt a maximum-likelihood approach to find our best fit solutions. Assuming a Gaussian line-of-sight velocity distribution, the log-likelihood of obtaining an empirical $v_{\rm rms}$ for one particular dynamical tracer $k$, given a set of model parameters $S =\{ r_s, \rho_s, \gamma, \Upsilon_*, \beta_k \}$ is \citep[e.g.,][]{Walker}:
\begin{equation}
\ln p_k(v_{{\rm rms_k}} | S_k) = - \frac{1}{2} \sum ^{N}_{i=1} \left[ 
\left( \frac{v_{{\rm rms}, i} - \sigma_p}{\Delta v_{{\rm rms}, i}} \right)^2 + \ln ( 2 \pi \Delta v_{{\rm rms}, i}^2)
\right]
\label{eq:likelihood}
\end{equation}
where $N$ is the sample size, whereas $v_{\rm rms}$ and $\Delta v_{\rm rms}$ are the empirical velocity dispersion profile of a given dynamical tracer and its uncertainty, respectively. 
The joint log-likelihood function that we want to minimize is therefore:
\begin{equation}
\log p(v_{\rm rms, k} | S) = \sum _{k=1} ^3 \log p_k(v_{{\rm rms}, k} | S_k) \\
\label{eq:likelihood2}
\end{equation}
with $S =\{ r_s, \rho_s, \gamma, \Upsilon_*, \beta_*, \beta_B, \beta_R \}$. 

Our model consists of seven free parameters: $r_s, \rho_s, \gamma, \Upsilon_*, \beta_*, \beta_B, \beta_R $. We point out that this dynamical model could be arbitrarily parametrized, folding in, for example, an orbital anisotropy varying with radius \citep[e.g.,][]{Wolf}. However, given the degenerate nature of the problem, we want to keep things simple and avoid to over-parametrized the model. 

To efficiently explore the parameter space, we use the \texttt{emcee} python module \citep{Foreman-Mackey}, which performs Markov-Chain Monte Carlo (MCMC) sampling of a parameter space, given a likelihood function (eq. \ref{eq:likelihood2}) and a prior function.

For the prior function, we adopt uniform priors over the following ranges: $\log r_s = \{ 1, 2 \}, \mbox{kpc}$ $\log \rho_s = \{ 5, 8 \} M_{\sun} \mbox{ kpc}^{-3}$, $- \log (1- \beta) =\{-1.5; 1 \}$, $\gamma =\{0,2 \}$ and $ \Upsilon_*= \{4 ,11.2 \} \Upsilon_{\sun, B}$. The prior function is defined such that the likelihood is forced to zero outside the above ranges. 

Our prior on $r_s$ is based on the values predicted for NFW haloes covering a wide range of galaxy masses: from $1\times 10^{11} M_{\sun}$ to $\approx 2 \times 10^{13} M_{\sun}$ \citep{Schaller14}. 
We will explore the effect of the chosen prior on $r_s$ further in \S \ref{sec:priors}.

The prior on the stellar mass-to-light ratio includes the values predicted by stellar population models for NGC 1407 \citep{Zhang,Humphrey06} in the case of a \citet{Kroupa01} IMF ($\Upsilon_{*,B} = 4-6.5 \Upsilon_{\sun, B}$) and a \citet{Salpeter55} IMF ($\Upsilon_{*,B} = 6-11.2 \Upsilon_{\sun, B}$). The large uncertainties on $\Upsilon_{*,B}$ come from systematics in the stellar population models.
 
The prior on $\gamma$ was chosen to be positive for consistency with similar studies \citep[e.g.,][]{Barnabe12,Cappellari13,Adams14} and because inner DM depression with $\gamma<0$ is a regime not yet explored by numerical simulations, with a few exceptions \citep[e.g.,][]{Destri14}.

We stress that the above priors were chosen to accommodate the values predicted by computer simulations, when available. This is common practice for dynamical modelling of stellar systems \citep[e.g.,][]{Cappellari13,Adams14}, and it forces the model parameters to stay in the physical regime. However, varying the chosen prior range will indeed affect the final results, although not strongly, as we will show in \S \ref{sec:priors}. 

The MCMC approach to data fitting is becoming increasingly popular for mass modelling of galaxies \citep{Mamon13,Agnello14B}, and in astrophysics in general. We found by experimentation that MCMC is strongly preferred over a classic ``gridded'' approach \citep[e.g.,][]{Schuberth12}, because it allows us to efficiently explore a much wider parameter space with reasonable computational time. 

\subsection{Characteristic DM parameters}
\label{sec:characteristic}

We use our MCMC results to infer the posterior distributions of some characteristic DM parameters. We compute the virial radius $r_{\rm vir}$, which is the radius where the mean density of the DM halo is $\Delta_{\rm vir}=101$ times the critical density $\rho_{\rm crit} =1.37 \times 10^{2} M_{\sun} \mbox{ kpc}^{-3}$ of the Universe. 
The virial mass and the concentration parameter are, respectively:
\begin{equation}
M_{\rm vir}=\frac{4}{3} \pi \Delta_{\rm vir} \rho_{\rm cri} r_{\rm vir}^3 \, \, \, \, \mbox{   and    } \, \, \, \, c_{\rm vir}=\frac{r_{\rm vir}}{r_s}.
\end{equation}

We also estimate the total mass-to-light ratio computed at the virial radius $(M_{\rm vir} /L)$, the DM fraction at $5 R_e$ $f_{\rm DM} (r=5 R_e) = M_{\rm DM}(r) / M(r)$, and the logarithmic radial gradient of the mass-to-light ratio \citep{Napolitano05}:
\begin{equation}
\nabla_l \Upsilon \equiv  \frac{R_e \Delta \Upsilon}{\Upsilon_* \Delta r}
\end{equation}
where $R_e$ is the stellar effective radius and the radial range $\Delta r$ was set to $\Delta r= 79$ kpc based on the radial extent of the fitted data points (see Figure \ref{fig:vrms}).

\subsection{Remarks on the data fitting}
\label{sec:remarks}

The substructures detected in the velocity dispersion profiles of stars and red GCs suggest that these features are real and not artefacts of the data. Interpreting the meaning of such features is beyond the scope of this paper. The $\sigma$-bumps in NGC~1407 could be the signature of major mergers \citep{PhoebeSchauer13}, or minor mergers \citep{Sharma11,Kafle}. In particular, \citet{PhoebeSchauer13} showed that the $\sigma$-bumps arise soon after the major merger involving at least one disc galaxy, and that this features are time invariant, meaning that the underlying galaxy has reached a state of dynamical equilibrium. More importantly, the $\sigma$-bumps show up only in the stellar and GC components, whereas the velocity dispersion profile of the DM halo is smooth after the merger.

Unlike the dynamical modelling with a \citet{Schwarzschild79} technique, the second-order Jeans equations used in our work are not designed to cope with kinematic substructures. In the case of non dynamical equilibrium, the generalized Jeans equations can in principle alleviate the effect of substructures on the mass modelling results \citep{Falco13}. On cluster scales, the impact of kinematic substructures on the inferred DM halo is negligible inside 2-3 virial radii \citep{FalcoNEW}. Also on galactic scales, the kinematic substructures have been shown to affect the final results up to 20 per cent \citep{Yencho06,Deason12,Kafle}.

\begin{figure*}
\includegraphics[scale=0.9]{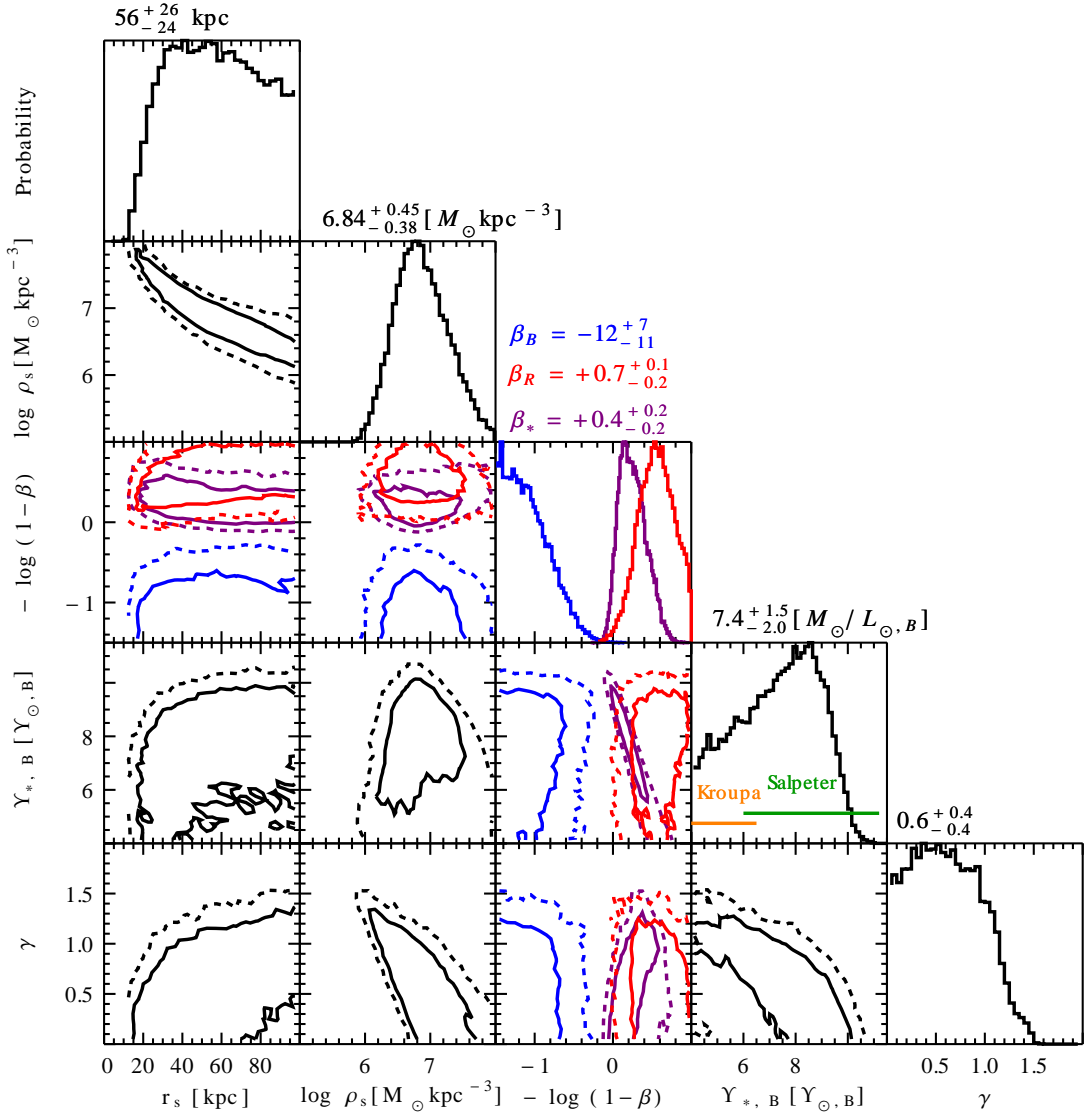} 
\caption{One and two dimensional projections of the posterior probability distributions of the seven free parameters of our model. Solid and dashed lines are the $1\sigma$ and $2\sigma$ contours of the distribution. The orbital anisotropy for the stars, blue GCs and red GCs is shown in purple, blue and red, respectively. The orange and green lines mark the expected value of $\Upsilon_*$ for a Kroupa IMF and Salpeter IMF, respectively. We show the linear scale radius $r_s$, although the fit was performed with the logarithmic equivalent ($\log r_s$). The median values of each distribution and relative $1\sigma$ uncertainty are quoted on each panel. Note that we report the actual value of $\beta$ and $r_s$, rather than $- \log (1-\beta)$ and $\log r_s$, respectively.}
\label{fig:triangle}
\end{figure*}

For our specific case, we will assess the effect of substructures on our results by manually masking the $\sigma-$bumps from the $v_{\rm rms}$ radial profile, as suggested by \citet{Kafle} and recently performed by \citet{Lane15}. In the following, we will focus on the results for the masked case only (i.e. we do not fit the open points and the stellar symbols in Figure \ref{fig:vrms}), and we will discuss the unmasked case in \S \ref{sec:priors}. 

The fit to the stellar data was performed using data points outside 2 arcsec because our model cannot reproduce the radial variation of orbital anisotropy found by \citet{Thomas14} within 2 arcsec. This also means that the central supermassive black hole \citep{Rusli2013a} is automatically excluded in our model, although the mass of the supermassive black hole is expected to have a minimal effect on the inferred DM halo.

\section{results}
\label{sec:results}

We run 40 MCMC chains (or walkers) with $5000$ steps each, which explore the parameter space simultaneously. The final acceptance rate of MCMC analysis for all our models is always between 35 and 40 per cent. We discard the first 15 per cent of the chains to account for the ``burn-in'' period, in which chains move from low likelihood regions to high likelihood regions in the parameter space. The modelling solutions and the best fits to the data are shown in Figure \ref{fig:triangle} and Figure \ref{fig:velfits}, respectively.

Figure \ref{fig:triangle} shows the 1D and 2D posterior distributions for our seven free parameters. For each parameter, we draw the contours containing 68 per cent ($1\sigma$) and 95 per cent ($2\sigma$) of the posterior distribution. The quoted best fit values and uncertainties in Figure \ref{fig:triangle} are the median of each distribution and the $1\sigma$ uncertainties (i.e., the 16th and 84th percentiles), respectively. 


Figure \ref{fig:triangle} shows that some of our model parameters are degenerate with each other. These degeneracies arise from the interplay between mass and anisotropy, and from the uncertain decomposition of the mass profile into stars and dark matter.  For example, there is a degeneracy between the slope $\gamma$ and the stellar mass-to-light ratio $\Upsilon_*$, such that large values of $\Upsilon_*$ imply smaller values of $\gamma$, and vice-versa \citep[see also][]{Newman13}.

Nonetheless, our joint maximum-likelihood analysis allows us to put broad constrains on both $\Upsilon_*$ and $\gamma$. The best fit stellar mass-to-light ratio in NGC~1407 is more consistent with a Salpeter-like IMF, although uncertainties propagate into the Kroupa regime. This result agrees within $1\sigma$ with the value $\Upsilon_{*,B}=6.6_{-0.8} ^{+0.9} \Upsilon_{\sun,B}$ found dynamically by \citet{Rusli2013b}, who assumed a logarithmic potential, and with $\Upsilon_{*,B}=7.6 \Upsilon_{\sun,B}$ found dynamically by \citet{Samurovic14}, who assumed an NFW halo. Note that both \citet{Rusli2013b} and \citet{Samurovic14} assumed a different parametrization of the DM halo with respect to our generalized NFW halo, as we discuss in \S \ref{sec:alternative}.

\begin{table*}
\centering
\label{mathmode}
\begin{tabular}{@{}l l l l l l l l l l}
\hline
Model & $\gamma$ & $\log M_{\rm vir}$ & $r_{\rm vir}$ & $M_{\rm vir}/L$ & $c_{\rm vir}$ &  $f_{\rm DM} (< 5 R_e)$ & $\nabla_l \Upsilon$ & $\Upsilon_*$ & $\log M_*$ \\
&  & [$M_{\sun}$] & [$kpc$] & [$M_{\sun} / L_{\sun, B}$] & & & & [$M_{\sun} / L_{\sun, B}$] & [$M_{\sun}$] \\ \hline
gNFW &  $0.6_{-0.4} ^{+0.4}$ & $13.34_{-0.21} ^{+0.22}$ & $724_{-109} ^{+135}$ & $260_{-100} ^{+174}$ & $13.2_{-3.5} ^{+6.7}$ & $0.83_{-0.04} ^{+0.04}$ & $1.0_{-0.2} ^{+0.3}$ & $7.4_{-2.0} ^{+1.5}$ & $11.79_{-0.13} ^{+0.08}$ \\
NFW  & $\equiv 1$				& $13.30_{-0.17} ^{+0.13}$  & $705_{-87} ^{+76}$ & $240_{-78} ^{+87}$ & $10.3_{-1.7} ^{+3.1}$ &  $0.85_{-0.04} ^{+0.03}$ & $1.1_{-0.3} ^{+0.4}$ & $6.1_{-1.2} ^{+1.2}$ & $11.71_{-0.10} ^{+0.08}$\\ 
Cored 	& $\equiv 0$			& $13.38_{-0.25} ^{+0.31}$  & $750_{-133} ^{+201}$ & $289_{-128} ^{+300}$ & $18.9_{-4.8} ^{+7.9}$ & $0.80_{-0.03} ^{+0.03}$ & $0.9_{-0.1} ^{+.0.2}$ & $8.8_{-1.4} ^{+0.9}$ & $11.87_{-0.07} ^{+0.04}$  \\ 
LOG  	& $ - $			&  $13.72_{-0.19} ^{+0.17}$  &  $974_{+133} ^{+136}$ & $632_{-226} ^{+303}$ & $-$ & $0.79_{-0.03} ^{+0.03}$ & $0.8_{-0.1} ^{+0.2}$ & $9.4_{-1.1} ^{+0.6}$ & $11.90_{-0.05} ^{+0.03}$ \\ 
LOG  	& $ - $ 			&  $13.51_{-0.12} ^{+0.11}$ & $826_{+73} ^{+78}$ & $385_{-93} ^{+119}$ & $-$ & $0.84_{-0.02} ^{+0.02}$ & $1.0_{-0.2} ^{+0.2}$ & $\equiv 6.6$ & $11.74$ \\ 
\hline
\multicolumn{10}{|c|}{prior: $10 < r_s < 250$ kpc} \\
\hline
gNFW & $0.7_{-0.5} ^{+0.4}$ & $13.47_{-0.26} ^{+0.31}$  & $801_{-146} ^{+216}$ & $352_{-159} ^{+369}$ & $9.6_{-4.0} ^{+7.0}$ &  $0.82_{-0.04} ^{+0.05}$ & $1.0_{-0.2} ^{+0.4}$ & $7.6_{-2.0} ^{+1.5}$ & $11.80_{-0.14} ^{+0.08}$ \\
\hline
\multicolumn{10}{|c|}{prior: $10 < r_s < 500$ kpc} \\
\hline
gNFW & $0.8_{-0.5} ^{+0.4}$ & $13.59_{-0.33} ^{+0.44}$  &  $880_{-201} ^{+361}$ & $466_{-251} ^{+842}$ & $8.1_{-3.9} ^{+7.7}$ & $0.82_{-0.04} ^{+0.05}$ & $1.0_{-0.2} ^{+0.4}$ & $7.8_{-2.0} ^{+1.4}$ & $11.82_{-0.13} ^{+0.07}$ \\
\hline
\end{tabular}
\caption{Solutions to the Jeans equations for different halo parametrizations. The quoted best fit values are the median of the posterior distributions of each parameter (defined in the text), with uncertainties representing the 16-th and 86-th percentile. Solutions are shown for a generalized NFW halo (gNFW), a standard NFW halo ($\gamma\equiv1$), a cored halo ($\gamma\equiv0$), and a logarithmic (LOG) halo, respectively. The last two columns give the results for a gNFW model when the prior on $r_s$ is relaxed to the quoted values.}
\label{tab:results} 
\end{table*}

The inner slope of the DM halo in NGC~1407 is found to be intermediate between a flat core and a cuspy (NFW) profile, and we cannot discriminate between these two regimes. A similar conclusion was also reached by \citet{NewmanB} in galaxy clusters. Our knowledge of $\gamma$ is therefore limited by the uncertainty on the stellar mass-to-light ratio: a Kroupa IMF ($\Upsilon_{*,B}<6$) implies cuspy DM haloes, whereas a Salpeter IMF implies cored DM haloes. Nevertheless, our results rule out, at the $ 2 \sigma$ level, a very cuspy inner DM slope, as found in M87 (with slope $\gamma\approx 1.6$) by \citet{Agnello14B}. The flattening of the DM density profile may be linked to many competing physical processes, such as baryonic feedback \citep{Governato12}, self-interacting DM \citep{Rocha13}, or dynamical friction from baryonic clumps \citep{El-Zant01}. AGN feedback, known to have occurred in NGC~1407 \citep{Giacintucci12}, can indeed produce central DM cores \citep{Cole11,DiCintio13,Martizzi13}. However, this would raise the question on why the modelling of M87, which also hosts an active AGN, supports a central DM cusp, rather than a DM core. 

The orbital anisotropy of the blue GCs is found to be very tangential $(\beta_B \approx -12)$, as first suggested by \citet{Romanowsky09}. This finding is in contrast to computer simulations which predict that halo particles reside on radial orbits \citep[e.g.,][]{Dekel05,Sommer-Larsen06}. On the other hand, the results for the stars $(\beta_* \approx +0.4)$ and red GCs $(\beta_R \approx +0.7)$ are indeed suggestive of radial orbits. This result supports the idea that stars and red GCs have a similar origin. We recall that the value of $\beta_*$ is the average orbital anisotropy outside the core radius (we only fit the stellar data outside 2 arcsec). This result also agrees with the results of \citet{Thomas14}, who found galaxies with central stellar cores to have radial anisotropies outside the core radius. Considering the strong deviation from isotropy found in our results, the assumption of Gaussianity in our model (\S \ref{sec:maximum}) could in principle bias our fits.

The posterior distributions for the characteristic DM parameters defined in \S \ref{sec:characteristic} are shown in Figure \ref{fig:par}. The best fit values are given in Table \ref{tab:results}. 
The results from this Table are discussed in Section \ref{sec:discussion}. 
We anticipate that our estimate of $M_{\rm vir}/L$ is consistent with the upper limit of $M_{\rm vir} / L \sim 300 M_{\sun}/ L_{\sun}$ found by \citet{Su14} (although the reader should see \S \ref{sec:discussion} for a full comparison with X-ray results), 
This result supports the suspicion that NGC~1407 is not at the centre of an extremely DM dominated group, but it is very sensitive to the assumed prior on $r_s$, as we will discuss in \S \ref{sec:priors}.

\begin{figure}
\includegraphics[width=\columnwidth]{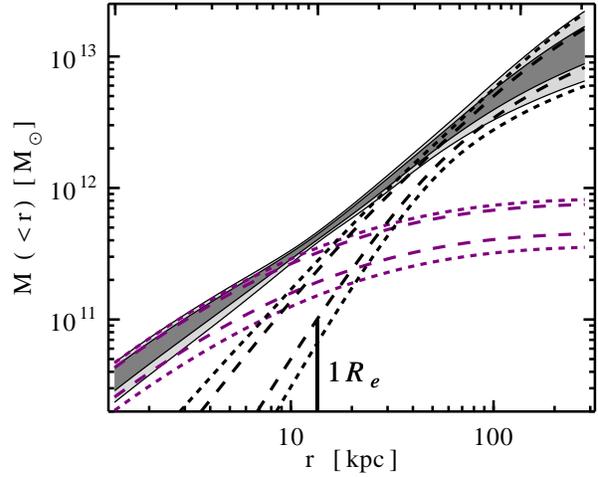} 
\caption{Best fit cumulative mass profile of NGC~1407. Dark grey and light grey contours show the $1\sigma$ and $2\sigma$ uncertainties on the total mass profile (i.e., the gNFW model in Table \ref{tab:results}). Black and purple lines show $1\sigma$ and $2\sigma$ envelopes of the DM component and of the stellar mass component, respectively. The uncertainties on the mass profile are minimized at the stellar effective radius $R_e\approx13$ kpc. }
\label{fig:massprofile}
\end{figure}

Lastly, we show the cumulative mass profile of NGC~1407 in Figure \ref{fig:massprofile}. The total mass profile was decomposed into DM mass $M_d (r)$ and stellar mass $M_*(r)$ to show the relative contributions of these two components. In Figure \ref{fig:massprofile} one can see that the stellar mass dominates within the stellar effective radius $R_e$, and also that the modelling uncertainties at this radius are minimized. We will further come back to this point in \S \ref{sec:pinch}.


\subsection{Alternative models}
\label{sec:alternative}

\begin{figure}
\includegraphics[scale=0.75]{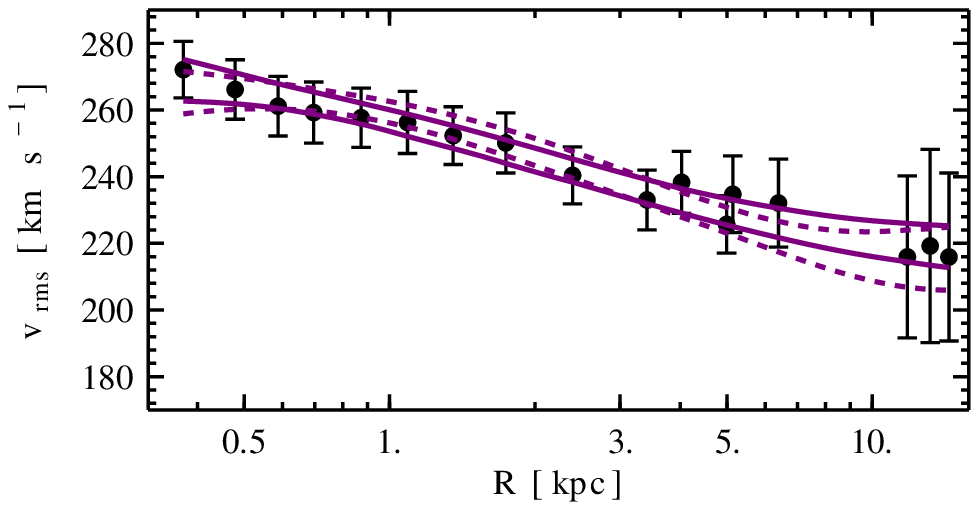} 
\includegraphics[scale=0.75]{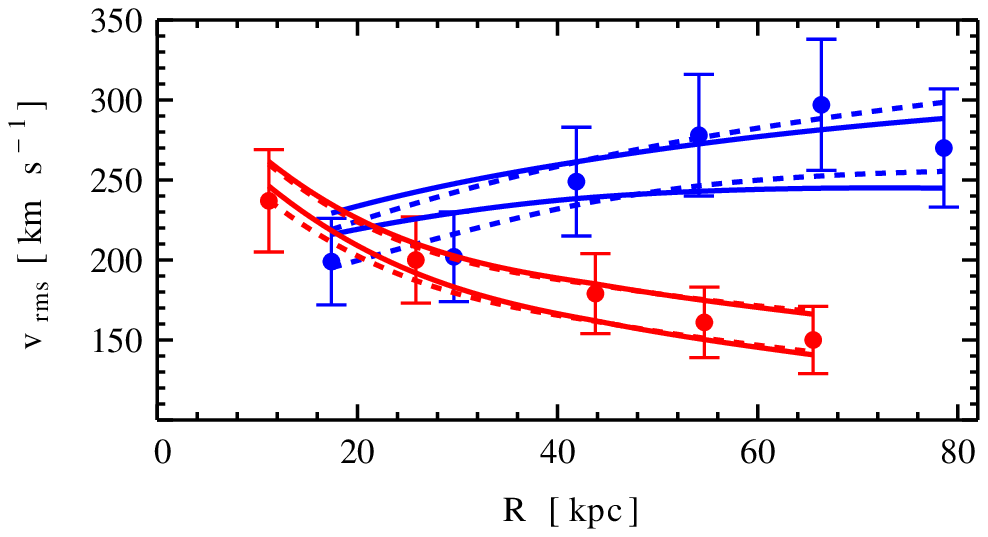} 
\caption{Best fit to the velocity dispersion profiles. Top and bottom panels show the $v_{\rm rms}$ profiles of stars (in log scale), and blue and red GCs, respectively. The thick and dashed lines are the $1\sigma$ envelope of the best fit $v_{\rm rms}$ profile for a gNFW and LOG model, respectively. The two models can fit the data equally well.} 
\label{fig:velfits}
\end{figure}

Our results cannot distinguish between DM cusps and cores in NGC~1407. Nonetheless, it is instructive to examine the outcome of our model when an NFW ($\gamma \equiv 1$) and a cored halo ($\gamma \equiv 0$) are assumed. By doing so, the NFW solutions can be compared both to $\Lambda$CDM predictions and to previous studies of NGC~1407 which adopted this form of parametrization (see Section \ref{sec:discussion}). Therefore, we fix $\gamma$ in eq. \ref{eq:Hern} to either 0 or 1, and solve the Jeans equations for the remaining six free parameters. 

\begin{figure}
\includegraphics[scale=0.75]{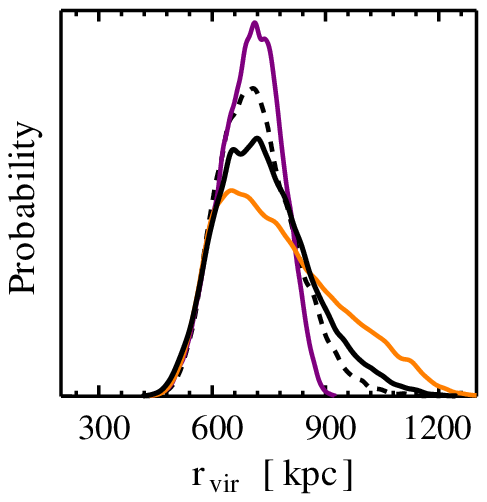} 
\includegraphics[scale=0.75]{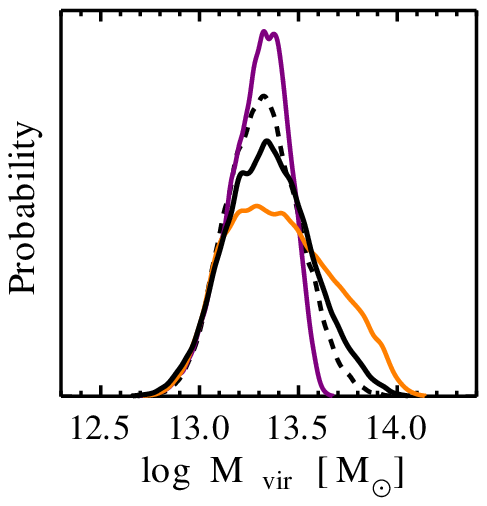} 
\includegraphics[scale=0.75]{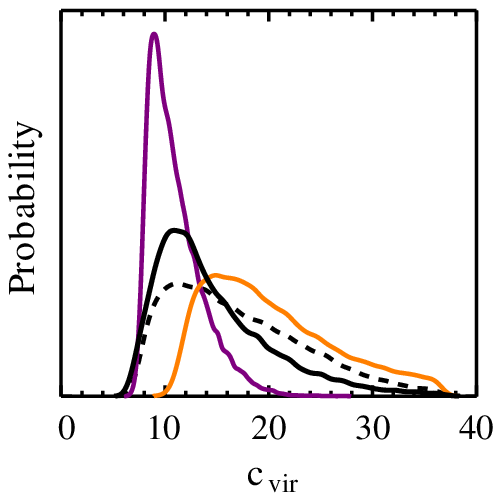} 
\includegraphics[scale=0.75]{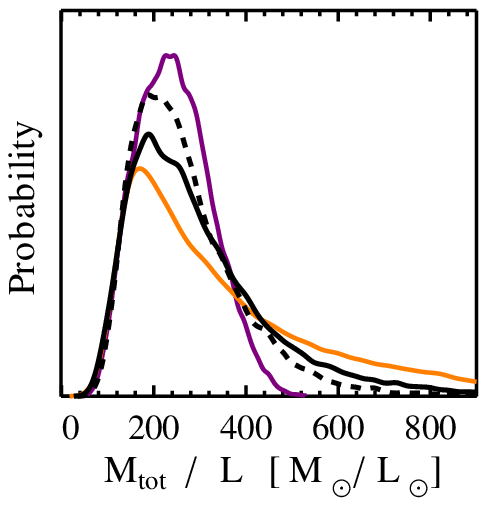} 
\caption{Characteristic DM parameters. Panels show the posterior distributions of the virial radius $r_{\rm vir}$, virial mass $M_{\rm vir}$, total mass-to-light ratio $M_{\rm vir}/L$ and concentration parameter $c_{\rm vir}$ respectively. The black and black dashed line corresponds to the solutions for a gNFW halo (i.e., $\gamma$ free to vary) when kinematic substructures are excluded or included in the fit, respectively. Purple and orange colours represent the posterior distributions for an NFW $(\gamma \equiv 1)$ and cored $(\gamma \equiv 0)$ halo, respectively.}
\label{fig:par}
\end{figure}

The DM parameters inferred from this exercise are given in Table \ref{tab:results} and shown in Figure \ref{fig:par}.  Generally speaking, the peaks of the NFW and cored distributions (purple and orange histograms in Figure \ref{fig:par}) are consistent with each other, and so are the respective best fit parameters. However, Figure \ref{fig:par} also shows that a cored halo requires more massive and extended DM haloes. Conversely, the posterior distributions of the NFW halo are fairly symmetric with smaller uncertainties. 

This difference between cored and cuspy solutions is related to the degeneracy between $\gamma$ and $\Upsilon_*$. Imposing a DM core means that more luminous matter can be added in the central regions, explaining the large values of $\Upsilon_*$, whereas more DM can be spread in the outer regions, explaining the large values of $M_{\rm vir}$ and $f_{\rm DM}$. As a consequence, the DM halo is more concentrated (larger $c_{\rm vir}$). Having a more accurate estimate of $\Upsilon_*$ would help to break this degeneracy. 

We now model the DM halo of NGC~1407 using an alternative parametrization of the DM distribution, known as the cored logarithmic (LOG) model \citep{Binney}:
\begin{equation}
M_d (r) = \frac{1}{G} \frac{v_0 ^2 r^3}{r_0 ^2 + r^2}
\label{eq:LOG}
\end{equation} 
where $r_0$ and $v_0$ are the core radius and the asymptotic circular velocity of the halo, respectively.
This model has six free parameters: $r_0, v_0, \Upsilon_*, \beta_*, \beta_B, \beta_R $.

Our maximum-likelihood analysis returns $r_0 = 44_{-17} ^{+19}$ kpc and $v_0 = 483_{-67} ^{+69} \kms$. The estimates of $\beta$ are similar to those in Figure \ref{fig:triangle}. The characteristic DM parameters for the LOG model are given in Table \ref{tab:results}. The results for the LOG model are more similar to those of the cored gNFW solutions, because both these models have a central DM core. However, the outer slope of the LOG profile ($r^{-2}$) allows for more DM in the outer regions compared to a gNFW, which declines instead as $r^{-3}$. The fit to the $v_{\rm rms}$ profiles of the three tracers is satisfactory. From Figure \ref{fig:velfits} it can be seen that both a LOG model and a gNFW model can fit the data equally well. 

\citet{Rusli2013b} also modelled the DM halo of NGC~1407 with a LOG potential. They found  $r_0 = 10.9$ kpc and $v_0 = 340 \kms$, which is only marginally consistent with our results. A caveat is that Rusli et al. used different data and a different modelling technique, which may bias the comparison with our results. Moreover, they focused on a baryon dominated region of the galaxy ($<1 R_e$), meaning that their estimate of $\Upsilon_{*,B}=6.6_{-0.8} ^{+0.9} \Upsilon_{\sun,B}$ should in principle be more accurate than our findings. Therefore, we fix $\Upsilon_*$ in our model to the best fit value from Rusli et al., and we study how this affects the inferred DM halo parameters.  

We find $r_0=17 \pm 4$ kpc and $v_0 = 409_{-37} ^{+39} \kms$, with the respective DM parameters listed in Table \ref{tab:results}. As expected, removing the dependency on $\Upsilon_*$ significantly reduces all modelling uncertainties, and it produces results more consistent with our gNFW solutions, but still marginally consistent with those of \citet{Rusli2013b}. In order for the LOG and gNFW models to be consistent with each other, the true value of $\Upsilon_*$ must be somewhere in the range $\Upsilon_{*,B}=5-8 \Upsilon_{\sun,B}$. This range is more skewed towards a Salpeter IMF according to stellar population models (see \S \ref{sec:maximum}).

\begin{figure*}
\includegraphics[scale=1.]{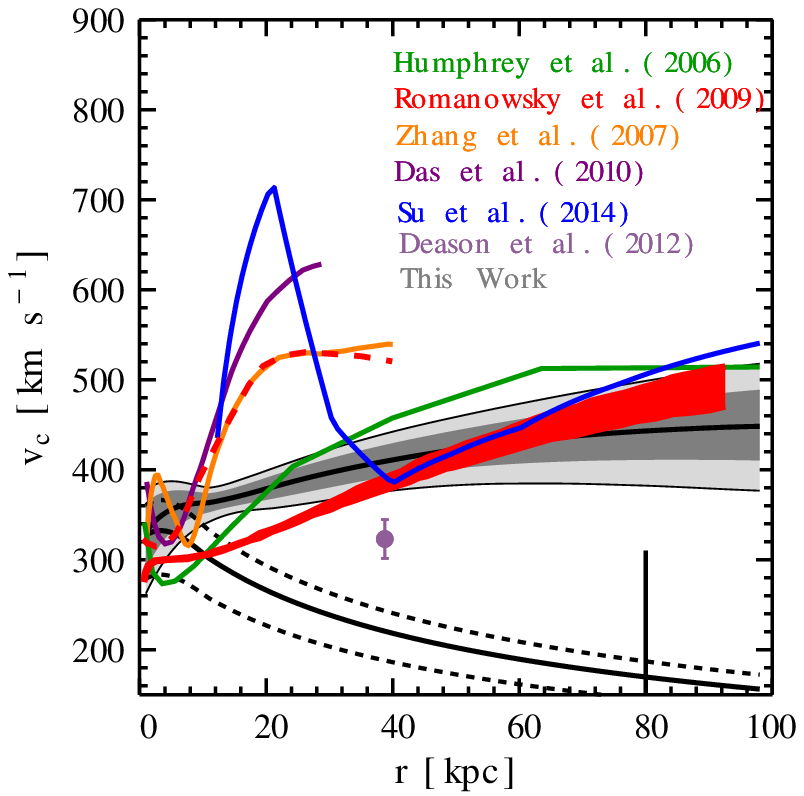} 
\includegraphics[scale=1.]{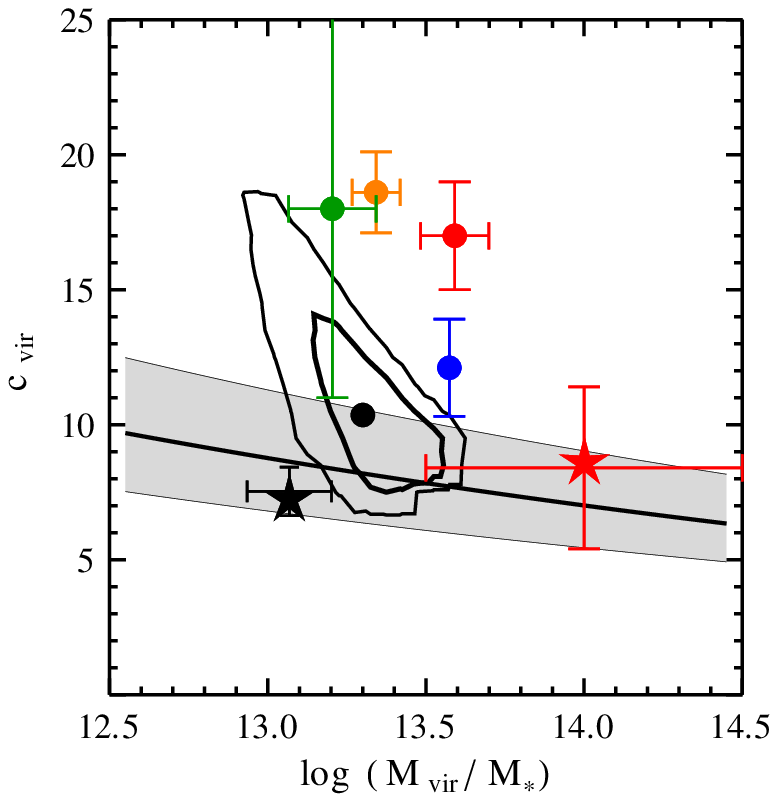} 
\caption{Comparison with literature studies and with $\Lambda$CDM predictions. \textit{Left Panel.} Circular velocity profiles, expressed as $v_c = (G M /r)^{1/2}$. The plot shows our best fit circular velocity curve with $1\sigma$ (grey) and $2\sigma$ (light grey) envelopes. The black dashed lines are the $1\sigma$ results for the stellar mass only. X-ray results are labelled and coloured accordingly. Uncertainties in most cases are not shown for clarity. The red dashed line correspond to the X-ray analysis of \citet{Romanowsky09}. The vertical black line marks the outermost data point used in our dynamical model.
\textit{Right Panel.} Virial mass - concentration relation. When an NFW halo is assumed, our results are shown as black contours ($1\sigma$ and $2\sigma$). Results from previous studies, when available, are colour coded as on the left panel. The red star and red circle correspond, respectively, to the GC-based and X-ray analysis of \citet{Romanowsky09}. The black star is the GC based result of \citet{Samurovic14}. The black line and $1\sigma$ scatter is the expected relation in a $\Lambda$CDM Universe (eq. \ref{eq:cvir}).}
\label{fig:comparison}
\end{figure*}

\section{Discussion}
\label{sec:discussion}

In this section we discuss our results in the light of previous mass estimates of NGC~1407 and in the context of $\Lambda$CDM predictions. Moreover, we want to address earlier findings that NGC~1407 might be at the centre of an extremely dark matter dominated group, and we want to compare our results to discrete mass estimators.

\subsection{Comparison with previous studies}

We compare our results with literature dynamical models of NGC~1407, with the caveat that these were obtained with different datasets and modelling techniques. The latter usually involved fixing $\Upsilon_*$, $\gamma$ and $\beta$, whereas we leave these parameters free to vary in our model.

The mass of NGC~1407 has been modelled via X-rays \citep{Humphrey06,Zhang,Romanowsky09,Das11,Su14}, GCs \citep{Romanowsky09,Deason12,Samurovic14} and stellar spectroscopy \citep{Saxton10,Rusli2013b,Thomas14}. Moreover, the dynamics of the dwarf galaxies around NGC~1407 suggest a total mass-to-light range $M_{\rm vir}/L\approx 300-2500 M_{\sun} / L_{\sun}$ and $M_{\rm vir} = 10^{13} - 10^{14} M_{\sun}$ \citep{Gould93,Quintana94,Tully05,Trentham06}. On the other hand, the dynamics of X-ray, stars, and GCs are suggestive of a normal DM content for this galaxy, with $M_{\rm vir}/L\approx300  M_{\sun} / L_{\sun}$. 

Figure \ref{fig:comparison} shows the comparison between our inferred mass profile and previous studies. The total mass profiles are shown in terms of circular velocity $v_c ^2 = G M /r$, where we recall that $M (r) = M_*(r) + M_d (r)$.

The hydrostatic equilibrium equation applied to the hot gas in NGC~1407 produces steeply rising mass profiles, along with a ``kink'' in the inner regions. These features have been detected in bright ellipticals \citep[e.g.,][]{Humphrey06,Napolitano14}, but they are not seen in our circular velocity curve in Figure \ref{fig:comparison}. The extreme behaviour of X-ray results may be linked to the X-ray gas not being in hydrostatic equilibrium \citep[e.g.,][]{Humphrey13}. For the specific case of NGC~1407, the non-equilibrium scenario is supported by disturbances in X-ray maps \citep{Su14}. In fact, the resulting $v_c (r)$ profile of \citet{Su14} agrees with our results, probably because they analyzed a relatively relaxed region of the X-ray gas outside 40 kpc. 
The results of \citet{Humphrey06} are marginally consistent with ours, but this is probably driven by their adopted stellar mass-to-light ratio $(\Upsilon_{*, B} = 2-3.6 \Upsilon_{\sun, B}$), which is a factor of 2 smaller than our fitted value $(\Upsilon_{*, B} = 7.4_{-2.0} ^{+1.5} \Upsilon_{\sun, B}$). 
Despite the disagreement between our results and X-ray results, the latter find normal mass-to-light ratios of $M_{\rm vir}/L\approx300 M_{\sun} / L_{\sun}$, in agreement with our results. 

The GC-based results of \citet{Romanowsky09} (their GI, GR and GT models) imply a very massive DM halo for NGC~1407 ($M_{\rm vir} =  3.4-26 \times 10^{13} M_{\sun}$ at a distance of 20.9 Mpc), whereas we find $M_{\rm vir} = 2.2_{-0.9} ^{+1.3} \times 10^{13} M_{\sun}$. Using a GC dataset very similar to ours, \citet{Samurovic14} inferred an NFW halo with $M_{\rm vir}= 1 \times 10^{13} M_{\sun}$, the lowest among the literature studies of NGC~1407, and a stellar mass-to-light ratio $\Upsilon_{*,B} = 7.6 \Upsilon_{\sun, B}$, which is in marginal agreement with our NFW solution. 

Our model solution may be characterized by the log--slope of the total density profile (stars + dark matter), $-\gamma_{\rm tot}$.
Inside 1~$R_e$, we find $\gamma_{\rm tot} \simeq 1.3$, which is intermediate to the gravitational lensing results for groups and clusters ($M_{200} \sim 10^{14}$ and $\sim10^{15} M_{\sun}$, respectively; \citealt{Newman15}). Over the range 1--4~$R_e$, the slope of $\gamma_{\rm tot} \simeq 1.4$ is much shallower than the $\gamma_{\rm tot} \sim 2.3$ found for fast-rotator early-type galaxies (Cappellari et al. 2015). This difference is not simply driven by a trend with stellar mass, as several of the fast rotators were of comparable mass to NGC~1407, but appears to reflect a distinction in the halo masses.

\citet{Deason12} used a power-law distribution function model for the dynamics of the total GC system, based on the data from \citet{Romanowsky09}. They found $v_c = 323 \pm 20\kms$ at 39 kpc, which is lower than all the other results at the $3 \sigma$ level (Figure \ref{fig:comparison}).  We suspect this is an effect of the restriction of those models to mass density slopes of isothermal and steeper ($\gamma_{\rm tot} \geq 2$). For reference, we estimate a DM fraction of $f_{\rm DM} = 0.83 \pm 0.04$ at $5 R_e$, whereas Deason et al. found $f_{\rm DM}\approx0.6 - 0.85$ depending on the adopted IMF. 

The trend emerging from the comparison with literature studies of NGC~1407 is that modelling results can be strongly biased depending on a number of factors, such as the type of dynamical tracer used, the modelling technique adopted and the assumed (or inferred) stellar mass-to-light ratio.

\subsection{Comparison with simulations}

We now discuss our results in the context of a $\Lambda$CDM cosmology, and compare our findings with properties of relaxed DM haloes at $z=0$. The $\Lambda$CDM model predicts a well defined $M_{\rm vir} - c_{\rm vir}$ relation \citep[e.g.,][]{Bullock01,Wechsler02,Schaller14}. Adopting a \textit{Planck} cosmology \citep{Planck14}, \citet{Dutton14} found:
\begin{equation}
c_{\rm vir} = 13.7  \left( \frac{M_{\rm vir}}{10^{11} M_{\sun}} \right)^{-0.097},
\label{eq:cvir}
\end{equation}
with a $1\sigma$ scatter of 0.11 dex at fixed $M_{\rm vir}$.

In order to compare our results with eq. \ref{eq:cvir}, we rely on our NFW solutions for a fair comparison. The right panel of Figure \ref{fig:comparison} shows that our best fit NFW halo is consistent with $\Lambda$CDM predictions within $1 \sigma$. On the other hand, the steeply rising $v_c(r)$ profiles from X-ray studies tend to produce large DM concentrations relative to $\Lambda$CDM predictions. The results of \citet{Su14} are in better agreement with our results, as expected. The large DM concentration inferred with X-ray modelling has been found in some \citep{Humphrey06,Ettori10,Oguri11}, but not all \citep{Gastaldello07}, X-ray datasets. However, at least in galaxy clusters, this discrepancy can be explained by invoking baryon physics \citep{Fedeli12,Rasia13}. 

For an NFW halo, the inferred logarithmic mass-to-light gradient is $\nabla_l \Upsilon = 1.1_{-0.3} ^{+0.4}$, which is consistent with the theoretical $M_* - \nabla_l \Upsilon$ relation from \citet{Napolitano05}. In conclusion, if we assume that the DM halo in NGC~1407 is an NFW halo, our results are fully consistent with $\Lambda$CDM predictions for a wide range of $\Upsilon_*$.

We cannot put strong constraints on the IMF, but we notice that our fitted stellar mass-to-light ratio is always $\Upsilon_{*} > 5 \Upsilon_{\sun,B}$ (at the $1\sigma$ level) regardless of the DM parametrizations. This result supports the range of $\Upsilon_*$ values predicted by a Salpeter IMF, although it is still consistent with a Kroupa IMF within $1\sigma$. A Salpeter IMF in NGC~1407 is supported by recent claims that the IMF is bottom-heavy in galaxies as massive as NGC~1407 \citep{Cappellari12,Conroy13,Tortora13MOND}.

\subsection{Comparison with mass estimators}

A number of analytic techniques have been proposed to infer galaxy masses. Their aim is to bypass time-consuming dynamical modelling, and determine the total galaxy mass from two observables: the projected effective radius of a dynamical tracer and a single measurement of its velocity dispersion. This provides the galaxy mass at one particular (3D) radius, dubbed ``sweet spot'' \citep{Churazov10} or ``pinch radius'' \citep{Agnello14A}. It was shown \citep{Wolf} that at the pinch radius, $r_{\star}$, the uncertainties from modelling assumptions are minimized. This fact was exploited to infer the presence of DM cusps and DM cores in dwarf galaxies \citep[e.g.,][]{Walker}.  

Table \ref{tab:estimators} lists a representative set mass of estimators (expressed in terms of circular velocity), as well as the location of the pinch radius proposed by different authors. Note that the velocity dispersion $v_{\rm rms}$ to be used in Table \ref{tab:estimators} can be either the velocity dispersion averaged within a certain radius $\langle v_{\rm rms} ^2 \rangle$, or the velocity dispersion measured \textit{at} one particular radius $v_{\rm rms} ^2 (R)$. 

\begin{table}
\centering
\label{mathmode}
\begin{tabular}{@{}l l l}
\hline
Author & $r_{\star}$ & $v_c ^2 (r_{\star})$ \\
       &       &   [$\kms$] \\
\hline 
\cite{WalkerB} & $1.0 \times R_e$ & $2.5 \, \langle v_{\rm rms} ^2 \, \rangle_{\infty}  $\\
\citet{Wolf}  & $1.3 \times R_e$ &$ 3 \, \langle v_{\rm rms} ^2 \, \rangle_{\infty} $\\
\citet{Cappellari13} & $1.0 \times R_e$ &$ 2.5 \, \langle v_{\rm rms} ^2 \, \rangle_{R_e}$\\
\citet{Amorisco11A} & $1.7 \times R_e$ &$3.4 \, \langle v_{\rm rms} ^2 \, \rangle_{R_e}$\\
\citet{Agnello14A} & $R_M$ & $K \, v_{\rm rms} ^2 (R_{\sigma})$ \\  
\hline
\end{tabular}
\caption{Mass estimators in terms of circular velocity $v_c ^2 = G M(r_{\star})/r_{\star}$. $r_{\star}$ indicates the location of the pinch radius, with $v_c$ being the circular velocity at that radius. The constants $R_M$, $K$ and $R_{\sigma}$ are dependent on the S\`ersic index of the dynamical tracer and were extrapolated from Table 1 of \citet{Agnello14A}.}
\label{tab:estimators} 
\end{table}

\begin{figure}
\includegraphics[width=\columnwidth]{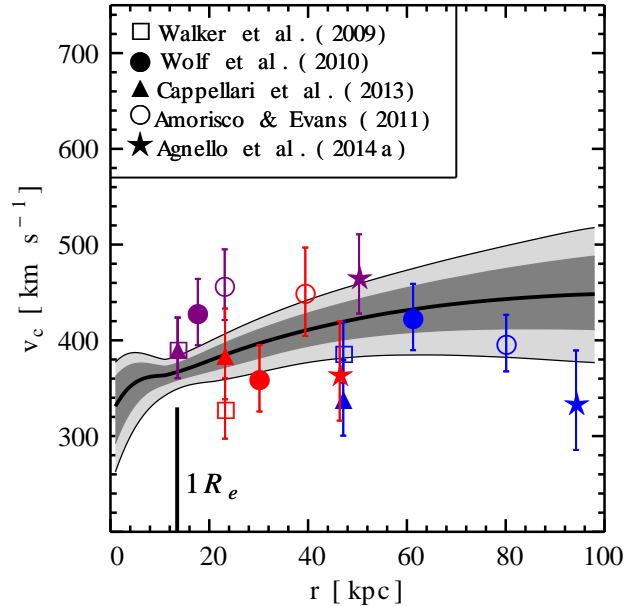} 
\caption{Comparison of mass estimators. The plot shows our best fit circular velocity curve with $1\sigma$ (grey) and $2\sigma$ (light grey) envelopes. Different symbols correspond to the mass estimators labelled on the top left, and defined in Table \ref{tab:estimators}. Purple, red and blue symbols correspond to mass estimators applied to stars, red GCs and blue GCs, respectively. Note that our modelling uncertainties are minimized at $1 R_e$. }
\label{fig:pinch}
\end{figure}

We insert the observables from stars, blue GCs and red GCs into the equations in Table \ref{tab:estimators} and derive discrete $v_c$ values for each tracer. We recall that the effective radii of stars, blue GCs and red GCs are 13.6 kpc, 47.1 kpc and 23.2 kpc, respectively.
The GC velocity dispersion averaged within a particular radius was calculated using eq. \ref{eq:sigma}. For the stars, we compute the luminosity-averaged velocity dispersion within a given radius. The velocity dispersion \textit{at} a given radius was inferred from the interpolated $v_{\rm rms}$ profiles in Figure \ref{fig:vrms}. The results are shown in Figure \ref{fig:pinch}.

Taken as a whole, the different mass estimators sample our circular velocity curve fairly well, and most estimators are consistent with our results within $2\sigma$. However, when taken singly, the mass estimators are not consistent with each other and they can return total masses which disagree up to 1.3 dex at a fixed radius.

This disagreement is not unexpected given that some mass estimators have classically been developed for dynamical modelling of dwarf galaxies, and have rarely been tested on large ellipticals, with some exceptions \citep{Cappellari13,Agnello14B}. Potential biases may rise from the significant velocity dispersion gradients we find in our data (see Figure \ref{fig:vrms}), in contrast to the flattish velocity dispersion profiles invoked by mass estimators for dwarf galaxies. However, the fact that mass estimators at large radii agree with our results is encouraging. Therefore, mass estimators should be used cautiously in elliptical galaxies with non-flat velocity dispersion profiles. 

\subsection{The pinch radius of NGC~1407}
\label{sec:pinch}

\begin{figure}
\includegraphics[width=\columnwidth]{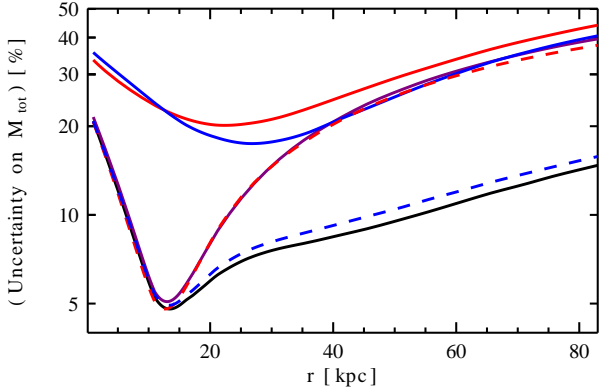} 
\caption{Percentage uncertainty on the inferred NGC~1407 total mass as a function of radius. Blue and red GCs are colour coded accordingly. Solid lines represent the uncertainty on $M$ when stars (purple) and GCs are modelled independently. The solid black line is the result for our \textit{joint} multi-tracer modelling. The blue dashed and red dashed lines are the results from the joint modelling of stars-blue GCs and stars-red GCs, respectively. The joint modelling of blue GCs and stars is enough to reduce uncertainties between 5 to 20 per cent in the range covered by the data.}
\label{fig:error}
\end{figure}

The cumulative uncertainties of our model results are minimized at 13.3 kpc, as shown by the black line in Figure \ref{fig:error}. This radius is in remarkable agreement with the projected stellar effective radius $1 R_e=13.6$ kpc, as also found by other authors \citep[e.g.,][]{Cappellari13,Walker,Agnello14A}. We find that at $1R_e$ the uncertainty is only 5 per cent, whereas it is 10 per cent between $0.5 R_e$ and $4 R_e$, and it is never larger than 20 per cent in the radial range covered by the data.  

The mass enclosed within $1 R_e$ is $M (< 1 R_e) = 4.2_{-0.2} ^{+0.1} \times 10^{11} M_{\sun}$. This result is valid for a large range of the orbital anisotropies, inner slopes of the DM profile and stellar-mass-to light ratios. The DM fraction is $f_{\rm DM} (1 R_e) = 0.37_{-0.13} ^{+0.13}$ per cent, which is similar to the values for high-mass galaxies in ATLAS$^{\rm 3D}$ \citep{Cappellari13}. Mass estimators at $1R_e$ give $M (< 1 R_e) = 4.8 \pm 0.7 \times 10^{11} M_{\sun}$ which is also in agreement with our results. This confirms that the projected stellar effective radius is the optimal radius to measure galaxy masses, albeit not the most useful for constraining DM.

It is interesting to quantify the effect of performing a joint multi-tracer dynamical modelling, relative to the results we would have obtained by modelling the dynamical tracers independently from each other. To do so, we perform dynamical modelling of NGC~1407 by using one dynamical tracer at a time (eq. \ref{eq:likelihood}). The results are shown as coloured solid lines in Figure \ref{fig:error}. We perform the same exercise by jointly modelling the pairs stars-red GCs and stars-blue GCs, respectively. These are shown as dashed lines in Figure \ref{fig:error}.

A clear picture emerges from Figure \ref{fig:error}. The modelling of GCs produces uncertainties larger than $\approx$20 per cent, whereas the modelling of the stars is able to constrain $M$ very accurately at $R_e$. Outside this radius, the estimate of $M$ from all three dynamical tracers is very loose. It is only by jointly modelling stars and GCs that the uncertainties are drastically reduced down to always $<20$ per cent in the radial range probed by the data. It is worth noting that the joint modelling of stars and blue GCs is enough to achieve satisfactory uncertainties, with the red GCs playing only a minor role. 

\subsection{Systematic uncertainties}
\label{sec:priors}

We investigate the effect of our modelling assumptions on the final results. The fact that the posterior distributions of some model parameters (for example $r_s$ in Figure \ref{fig:triangle}) tend to overfill the allowed parameter space is indeed a concern, although very common among similar studies \citep[e.g.,][]{Barnabe12,Kafle14}. This is driven by a set of degeneracies which prevent us from determining some parameters with satisfactory accuracy. 

A common practice is to choose uniform priors based on results from computer simulations, which is also our adopted strategy. This ensures that the model parameters stay in the physical regime. However, it is instructive to investigate how the choice of the priors affects our results. 

We focus on the scale radius $r_s$ only, leaving the analysis of the full parameter space as a future exercise. We recall that our analysis was carried out with the logarithmic prior $10< r_s < 100$ kpc. We relax the priors on $r_s$ to $10 <r_s < 250$ kpc and $10 <r_s < 500$ kpc respectively. The remaining free parameters were left unchanged. The results are listed in Table \ref{tab:results} for the gNFW model only. 

One can see that main effect of expanding the prior on $r_s$ is to boost the uncertainties on the DM parameters towards more massive and more extended DM haloes. The stellar mass is insensitive to our new prior choice because the stellar mass contribution is negligible at large radii. One should also bear in mind that $r_s$ cannot be constrained if this is larger than the radial extent of our data ($\approx 80$ kpc). For reference, we find $r_s \approx 50$ kpc.

We tested the effect of the adopted galaxy distance on the final results by modelling the galaxy at a distance of 20.9 Mpc (as in \citealt{Romanowsky09}), instead of the 28.05 Mpc used throughout this paper (see \S \ref{sec:distance}). We find that the distance has little impact (always within 1$\sigma$) on the characteristic DM parameters, total mass-to-light ratio and DM fraction. On the other hand, a closer distance returns a stellar mass-to-light ratio (and therefore a stellar mass) that is a factor of 1.3 smaller, $\Upsilon_* = 5.8_{-1.1} ^{+0.9} \Upsilon_{\sun, B}$, and a concentration parameter that is a factor of 1.2 larger $c_{\rm vir} = 15.8_{-4.9} ^{+7.9}$.

Our analysis was carried out by masking the kinematic substructures detected for the stars and for the red GCs, for the reasons explained in \S \ref{sec:remarks}. Figure \ref{fig:par} shows by how much some DM parameters vary when the kinematic substructures are included in the fit. The difference between the masked and unmasked models is small (well within the $1\sigma$ errors) and the impact on the results discussed in this paper is negligible.  The smaller uncertainties found from unmasked data is due to the inability of the Jeans equation to fit the kinematic substructures as discussed in \S \ref{sec:remarks}.

\section{Summary}
\label{sec:summary}

We have conducted in-depth dynamical modelling of the elliptical galaxy NGC~1407. 
Our approach consisted of solving, simultaneously, the spherical Jeans equations for three independent dynamical tracers in NGC~1407: galaxy stars within one effective radius $(1 R_e)$, and blue and red GCs out to $10 R_e$. This technique alleviates well-known model degeneracies, lowering the final uncertainties. 

Stellar data were constructed by combining long-slit spectroscopy with multi-slit spectroscopy from the Keck/DEIMOS multi-object spectrograph. We also analysed 153 blue GCs and 148 red GCs obtained from DEIMOS. Both stellar and GC data are products of the SLUGGS survey. 
The DM was parametrized with a generalized Hernquist profile, with an asymptotic outer slope dictated by $\Lambda$CDM predictions. The DM inner slope $\gamma$ was left free to vary in the attempt to discern whether the DM in NGC~1407 has a central cusp or a core (the cusp/core problem). The stellar mass-to-light ratio was also left free to vary within priors imposed by single stellar population (SSP) modelling. We used a Markov-Chain Monte Carlo method to explore the wide 7-dimensional parameter space. 

Although we cannot discriminate between a cored or cuspy DM profile in NGC 1407, our results are suggestive of a DM central slope shallower than a cosmological Navarro-Frenk-White (NFW). This result disfavours a steepening of the DM profile due to adiabatic contraction, and it may favour a scenario in which DM is evacuated from the galaxy centre via some physical mechanism, such as baryonic feedback or self interacting DM. Our knowledge of $\gamma$ is limited by the uncertainties on the mass-to-light ratio $\Upsilon_*$. Our estimate of $\Upsilon_*$ favours a Salpeter IMF,  supporting recent claims that the IMF becomes bottom-heavy in more massive galaxies. Our results are accurate between 5 to 20 per cent in the radial range probed by the data, which is a big improvement compared to dynamical models with single dynamical tracers.
We find that our modelling uncertainties are minimized at the stellar effective radius, establishing it to be the best radius for measuring galaxy masses.

We confirm that NGC~1407 is surrounded by a DM halo with an inferred total mass-to-light ratio of $M_{\rm vir}/ L = 260_{-100} ^{+174} M_{\sun} / L_{\sun}$. However, the value of $M_{\rm vir}/ L$ increases up to a factor of three if the prior on the DM scale radius is relaxed to $10< r_s < 500$ kpc. Therefore, we cannot determine whether or not NGC~1407 is at the centre of an extremely DM dominated group as claimed in previous studies.

We tested that the disagreement between our results and X-rays results is due to the X-ray gas not being in hydrostatic equilibrium in the inner regions of this galaxy. When a relatively relaxed part of the galaxy is considered, our findings are in better agreement with X-ray results.
When the DM halo of NGC~1407 is assumed to follow a cosmological NFW profile, our results are consistent with the predicted virial mass--concentration relation from $\Lambda$CDM. We compared our mass profile of NGC~1407 with a set of discrete mass estimators, finding a marginal agreement. However, we argue that mass estimators should be used cautiously in galaxies with strong velocity dispersion gradients because their results may be strongly biased.

\section*{acknowledgements}

We thank the referee for helpful and expeditious comments.
We thank Alister Graham, Zachary Jennings, Adriano Agnello and Chris Flynn for constructive discussions. This work was supported by NSF grant AST-1211995. We thanks the ARC for financial support via DP130100388. This research has made use of the NASA/IPAC Extragalactic Database (NED) which is operated by the Jet Propulsion Laboratory, California Institute of Technology, under contract with the National Aeronautics and Space Administration.  
Some of the data presented herein were obtained at the W. M. Keck Observatory, operated as a scientific partnership among the California Institute of Technology, the University of California and the National Aeronautics and Space Administration, and made possible by the generous financial support of the W. M. Keck Foundation. The authors wish to recognise and acknowledge the very significant cultural role and reverence that the summit of Mauna Kea has always had within the indigenous Hawaiian community. The analysis pipeline used to reduce the DEIMOS data was developed at UC Berkeley with support from NSF grant AST-0071048. The authors acknowledge the data analysis facilities provided by IRAF, which is distributed by the National Optical Astronomy Observatories and operated by AURA, Inc., under cooperative agreement with the National Science Foundation.

\bibliographystyle{mn2e}
\bibliography{ref}
\end{document}